# INERTIA EMULATION CONTRIBUTION OF FRADES 2 VARIABLE SPEED PUMP-TURBINE TO POWER NETWORK STABILITY

C. Nicolet, A. Béguin, M. Dreyer, S. Alligné, A. Jung, D. Cordeiro, C. Moreira

Abstract: This paper is addressing the quantification and the comparison of pumped storage power plants, PSPP, contribution to synchronous inertia and synthetic inertia when fixed speed and variable speed motor-generators technologies are considered, respectively. Therefore, a grid stability study was conducted by means of 1D SIMSEN simulation for the 2 x 395 MW PSPP Frades 2 in Portugal with both fixed speed and variable speed technologies in case of operation connected to an infinite power network or to an islanded 4.4 GW synchronous power network.

## 1 Introduction

The overarching objective of the XFLEX HYDRO H2020 European Project is to develop and demonstrate new technological solutions capable to improve efficiency and performance of Hydro Power Plants (HPP) with respect to the provision of several Electric Power Systems (EPS) services. Such developments are expected to actively contribute to the decarbonization of the European EPS, allowing HPP to increase their capabilities to provide advanced grid services to integrate increasing share of volatile renewable energy sources such as wind and solar, [1]. In this context, variable speed pumped storage power plants can play a strong role by supplying electrical power systems with a variety of ancillary services thanks to their operating flexibility, fast active power response time and large energy storage capability.

As far as the Continental Europe (CE) synchronous area (SA) is concerned, the following EPS frequency control ancillary services have been identified, [2]:
- Synchronous inertia;
- Synthetic inertia with response time < 0.5 s;
- Fast Frequency Response (FFR) with response time < 0.5 s - 2 s;
- Frequency Containment Reserve (FCR) with response time < 30 s;
- Automatic Frequency Restoration Reserve (aFRR) with response time in the range 30 s - 5 min, being automatically deployed by the system operator at the control centre;
- Manual Frequency Restoration Reserve (mFRR) achievable in less than 15 min, with a maximum Full Activation Time (FAT) of 12.5 min, being manually activated by the system operator;
- Replacement Reserve (RR) to be made available in less than 30 min.

The contribution of hydro power plants to the FFR, FCR, aFRR, mFRR and RR ancillary services can be evaluated using grid codes edited by Transmission System Operators (TSO) whenever applicable, see [3]. However, the evaluation of synchronous and synthetic inertia contributions requires a special attention since there is no grid code nor metrics available for these particular ancillary services.

This paper is addressing the quantification and the comparison of pumped storage power plants contribution to synchronous inertia and synthetic inertia when fixed speed and variable speed motor-generators technologies are considered, respectively. First, an analytical expression is derived to quantify the magnitude of the active power that a synchronous machine would inject or absorb in the case of a large power network frequency deviation. This simple analytical expression enables to show that the magnitude of active power injected or absorbed is proportional to the hydro unit mechanical time constant τm for a given Rate of Change of the Frequency (RoCoF).

To demonstrate the validity of the analytical approach, the XFLEX HYDRO demonstrator of Frades 2 Pumped Storage Power Plant (PSPP) located in Portugal, was considered. This PSPP operated under a maximum gross head of 431.8 mWC includes an upper reservoir, a headrace tunnel, a headrace surge tank, a penstock supplying two variable speed reversible Francis pump-turbines of 395 MW each equipped with Doubly Fed Induction Machines (DFIM) a tailrace surge tank, a tailrace tunnel and a lower reservoir.

Unlike synchronous machines, the rotational speed of variable speed units is decoupled from the power network frequency. Nevertheless, inertial response can be emulated by variable speed unit, using an appropriate power control structure to provide so-called synthetic inertia to the power network. Inertia emulation control structure is based on the swing equation in order to achieve active power injection or absorption which is also proportional to the RoCof like for a synchronous machine, and thus replicate the corresponding flywheel effect. To illustrate this, a specific grid stability study was conducted by means of 1D SIMSEN simulation for the demonstrator of Frades 2 with both fixed speed and variable speed technologies in case of operation connected to an infinite power network or to an islanded 4.4 GW synchronous power network.

## 2  Synchronous and synthetic inertia

### 2.1  Synchronous inertia

#### 2.1.1  Theoretical background

Synchronous inertia is the inherent capability of rotating synchronous machines directly connected to the power grid to store or inject their kinetic energy, which compensates for the frequency transient behaviour in the moments after an active power imbalance. A simplified power grid schematics with synchronous generators and loads and the transport system is shown in Fig. 1, illustrating active power and rotor speed typical waveforms before, during and just after a sudden load change. During the sudden load change, active power at all stators also changes quasi instantaneously. This is a fast transient grid phenomenon which can be considered instantaneous in front of the other transient regimes studied here. The magnitude of change of active power at each generator stator depend mainly on its position with respect to the load change and it is different for each generator. The load change has created a mismatch between the electro-mechanical torque ($T_{em,k}$) and the prime movers mechanical torque ($T_{mec,k}$) on each generators. This torque unbalance makes the speed deviates from the synchronous speed. Each speed deviation is different for each generator. However, thanks to the inherent synchronization capabilities of the generators, the speed oscillates and these oscillations eventually damp out. The average speed is the grid frequency. The frequency deviations described here will be limited by the FCR contributions of each generator which FCR service is enabled.

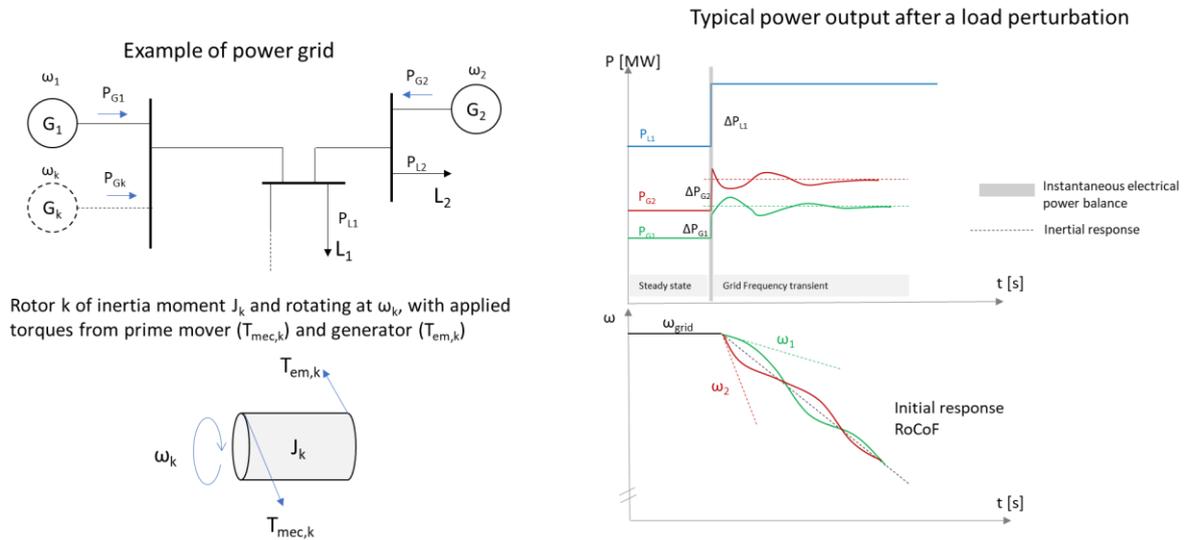

Fig. 1. Typical power and speed waveforms after a load change in a synchronous power grid.

It is intuitive that the higher the moment of inertia J of a rotor, the smaller is the acceleration for a given torque, thus limiting the speed deviation, hence the grid frequency. This is also expressed by the swing equations of a rotating mass. However, the moment of inertia itself ($J = M \cdot R^2$ in kgm$^2$) is not relevant enough, as it has to be associated with the power that the generator can deliver to the grid, *i.e.,* the rated power and the kinetical energy it is normally storing at nominal speed. This is known as the mechanical time constant as defined in equation (1). It is a measure of how fast the rotor or shaft can be accelerated to its nominal speed under the constant rated torque.

$$\tau_m = \frac{J \cdot \omega_{rated}^2}{P_{rated}} \quad (s) \tag{1}$$

The well-known swing equation, or motion equation of the rotor, can also be expressed with speed and torque expressed in per unit (symbol topped with a bar), see equation (2):

$$\tau_m \frac{d\overline{\omega}_k}{dt} = \overline{T}_{mec} - \overline{T}_{em} \tag{2}$$

The frequency derivative immediately following a sudden active power imbalance $\overline{\Delta P}$ *in per unit* on the grid is given by equation (3), see [4]:

$$\frac{d\omega_{grid}}{dt} = \frac{\overline{\Delta P}}{\tau_m} \omega_{rated} \tag{3}$$

The total mechanical time constant of the power grid $\tau_{m,sys}$, which accounts for all individual inertia synchronously connected to the grid, is defined by equation (4):

$$\tau_{m,sys} = \frac{\sum_k \tau_{m,k} \cdot P_{rated,k}}{\sum_k P_{rated,k}} \tag{4}$$

As seen here, the inertia constant (or the mechanical time constant) is the unique property that supports the grid frequency initial deviation. Hence, in the evaluation of the synchronous inertia as an ancillary service of synchronous generators, this is directly the inertia constant that is relevant for the scoring of the tested unit.

### 2.1.2 Instantaneous power contribution of synchronous inertia

From the equation (2) and (3) and neglecting speed deviation in front of rated speed, it can be shown that the instantaneous inertial response of synchronous generators, *i.e.,* power output without the oscillations due to inter generator synchronization, is proportional to the grid frequency derivative, as equation (5) shows:

$$\overline{\Delta P}_{Gk,inertial} = -\tau_m \frac{d\overline{\omega}_{grid}}{dt} \overline{\omega}_{grid} \approx -\tau_m \frac{d\overline{\omega}_{grid}}{dt} \text{ (for small frequency variation in pu)} \tag{5}$$

This expression of the instantaneous inertial response is used as a base for the design of a synthetic inertia controller, as presented in section 2.2.

### 2.1.3 Mean power contribution of synchronous inertia

Following an active power imbalance and assuming that the power network frequency goes from an initial angular speed value $\omega_1$ to $\omega_2$, one can express the kinetic energy injected or absorbed by the rotor, neglecting losses in shaft, generator winding and grid transmission system, with the equation (6):

$$\Delta E_{k,syn} = \frac{1}{2} J_k \left( \omega_1^2 - \omega_2^2 \right) \tag{6}$$

Moreover, the mean power variation injected or absorbed during the time $\Delta t$, for which the angular speed vary from $\omega_1$ to $\omega_2$ can be evaluated by equation (7):

$$\Delta P_{mean} = \frac{1}{2} \frac{J_k \left( \omega_1^2 - \omega_2^2 \right)}{\Delta t} \longrightarrow \frac{\Delta P_{mean}}{P_n} = \frac{1}{2} \frac{\tau_m}{\Delta t} \left( 1 - \left( \frac{\omega_2}{\omega_1} \right)^2 \right) \tag{7}$$

The equation (7) is used to quantify active power injection or absorption for different values of positive *RoCoF*, total rated power and mechanical time constants and are provided in the Table 1 considering the test case of the Frades 2 PSPP equipped with two reversible Francis pump-turbines of 395 MW rated power and described in chapter 3. This analytical approach shows that units operating the unit in turbine or in pump mode contribute both to synchronous inertia since both feature rotating masses storing angular kinetic energy similarly to a flywheel. It also shows that, for hydraulic short circuit operation (HSC) involving separated pump and turbine, the contribution to synchronous inertia is the sum of the pump and the turbine units synchronous inertia contributions. Finally, it shows that the active power injection or absorption is almost proportional to the *RoCoF*, at least for small network frequency deviation, and it is proportional to the mechanical time constant of the unit.

| PSPP | Σ|Pn| | RoCoF | Tm | ΔP/Pn | ΔP |
|---|---|---|---|---|---|
| Frades2 U1 | 395 MW | 0.5 Hz/s | 7.9 s | 8 % | 31 MW |
| Frades2 U1 (ref.) | 395 MW | 1 Hz/s | 7.9 s | 16 % | 63 MW |
| Frades2 U1 | 395 MW | 2 Hz/s | 7.9 s | 32% | 127 MW |
| Frades2 U1 | 395 MW | 2 Hz/s | 3.95 s | 16 % | 63MW |
| Frades2 HSC U1+U2 | 790 MW | 2 Hz/s | 7.9 s | 31% | 254 MW |

Table 1. Evaluation of synchronous inertia contribution for different values of RoCof, inertia and number units (HSC: Hydraulic Short-Circuit)) over a duration of Δt = 1 s using equation (7) for positive RoCoF and $\omega_1$=314.5 rad/s.

## 2.2 Synthetic inertia

Power electronic-interfaced energy sources do not change their power output in the moments after an active power imbalance on the grid, as illustrated in Fig. 2. This is because their control is on purpose maintaining constant output power according to the power setpoint. For these power sources to provide short-term frequency support like synchronous generators do, a proper control of the coupling interface is required, [5], [6]. While not providing pure synchronous inertia, they are able to swiftly adapt power output, driven by their control system, to deliver "synthetic inertia", provided some energy buffer is available within the primary energy source as it is the case for the kinetic energy stored in the rotating masses of wind or hydro facilities.

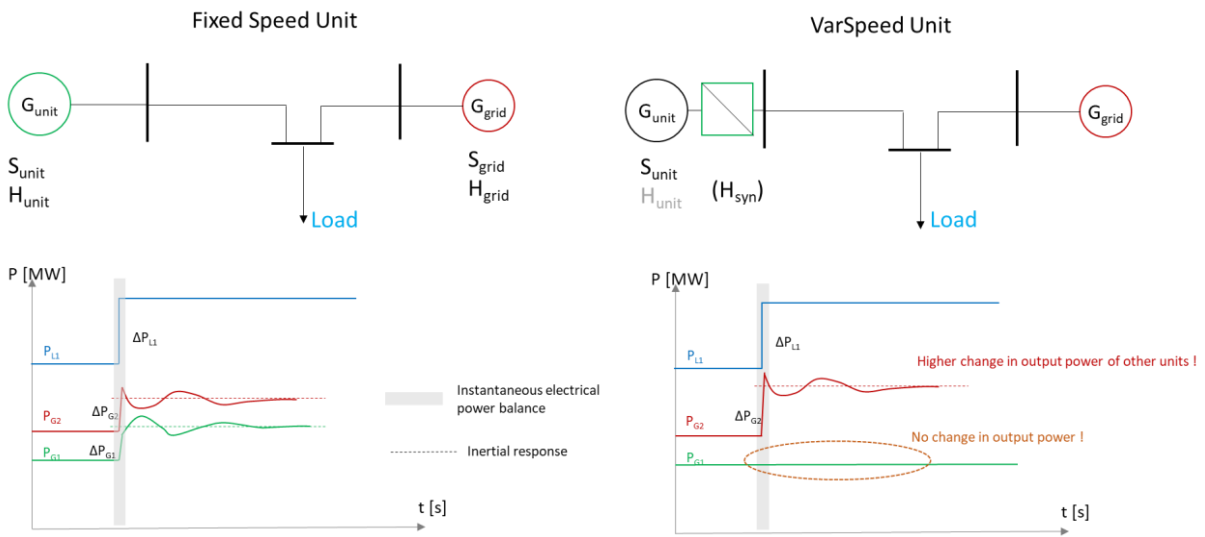

Fig. 2. Typical power and speed waveforms after a load change in a mixed power grid.

One possible function to emulate synchronous inertial response directly comes from the expression of synchronous inertial response, corresponding to equation (5). The transfer function is shown in the lower branch of Fig. 6. Its output is proportional to the filtered frequency derivative, with $K_d$ the derivative gain and $\tau_d$ the filter time constant, see [7], [8].

## 3  Presentation of Frades2 PSPP

### 3.1  PSPP description

Frades 2 hydroelectric plant is a PSPP built between 2010 and 2017 on the Rabagão river in the North of Portugal. The plant is composed of two high head, variable speed units made of two reversible pump turbines, coupled with 420 MVA motor-generator of the Doubly Fed Induction Machine, DFIM, technology [9]. The main hydraulic and power generation characteristics are given in table 2. The waterway includes a headrace tunnel, an upper surge tank followed by a sandtrap, a penstock that feeds the distributor of each unit, a lower surge tank and the tailrace tunnel, as shown in Fig. 3, a. The pump-turbine, which is illustrated in Fig. 3, b, is characterized by a specific speed of Nq=38 and a unit mechanical time constant of $\tau_m$= 7.9 s.

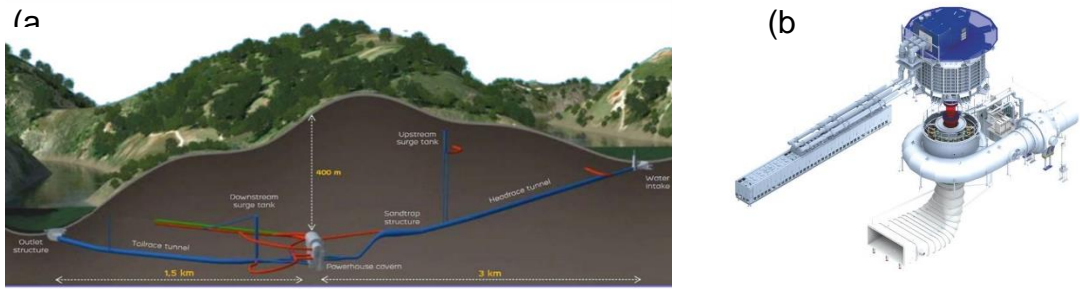

Fig. 3. Frades 2 pump storage power plant layout (a) and 3D view of the variable speed pump-turbine of Frades 2 PSPP (b).

| Francis pump-turbine | |
|---|---|
| Type | Francis type single-stage reversible pump-turbine |
| Head | Maximum 431.80 m, Minimum 413.64 m |
| Number of units & unit size | 2 units, 4.500 m |
| Turbine rotational speed range | 350 min$^{-1}$, 381 min$^{-1}$ |
| Mechanical power | Generating mode: 400 MW, 390 MW, 190 MW |
|  | Pumping mode: -300 MW, -381 MW, -390 MW |
| Rated mechanical power | 395 MW |
| Specific speed number | 38 SI |
| Mechanical time constant | 7.9 s |
| Motor-Generator | |
| Type of power generator | Asynchronous machine |
| Variable speed technology | DFIM |
| Rated apparent power | 420 MVA |
| Network frequency | 50 Hz |

Table 2. Frades 2 pump storage power plant characteristics.

## 3.2 Frades 2 PSPP modelling

To assess the compliance level of the DFIM technology against the set of ancillary services considered and to quantify corresponding performances, the Frades 2 PSPP has been modelled with the 1D SIMSEN software, see [10], [11], and validated [12]. The model, which is illustrated in Fig. 4, includes all the waterways and the two reversible Francis pump-turbines with their rotating mass inertia. The behavior of the reversible Francis pump-turbine is modelled with the 4-quadrants characteristics provided by the turbine manufacturer.

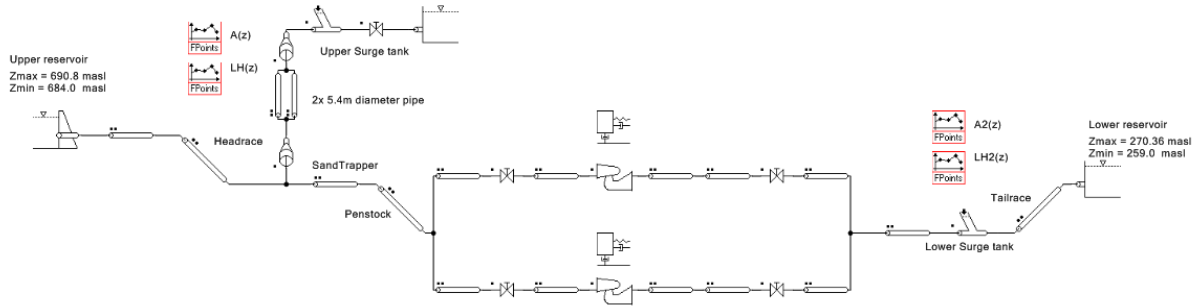

Fig. 4. SIMSEN model of the Frades 2 PSPP.

The variable speed technology allows to manage the power exchange with the grid with more flexibility than a conventional synchronous machine. In fact, two control strategies can be implemented to manage this power exchange, see [13], [14], [15], [16], [17], [18], [19], [20], [21], [22]. The first control strategy acts on the pump-turbine (PT) guide vane opening to regulate the power, while the power electronics of the motor-generator (MG) manage the speed of the unit. Due to the different dynamic time constants between electrical and hydraulic parts, this strategy is preferred to ensure a desired speed response, but it can result in a large deviation between the power output of the unit and the set point during transients. The second control strategy uses the power electronics to regulate the output power by managing the active power, while the rotational speed is controlled by the unit speed governor which manages the opening of the guide vanes, as schematized in Fig. 5. This strategy is the main one used in turbine operation as it allows a fast power response of the unit to an active power setpoint change, while the speed can be adjusted more slowly by the guide vane to achieve optimal efficiency speed. However, depending on the requested change of the unit load, the unit rotational speed may exceed the allowable operating range, [12]. In that case, it is necessary to operate a strategy switch by taking advantage of the fast active power response of the power electronics to preserve the unit from over/under speed.

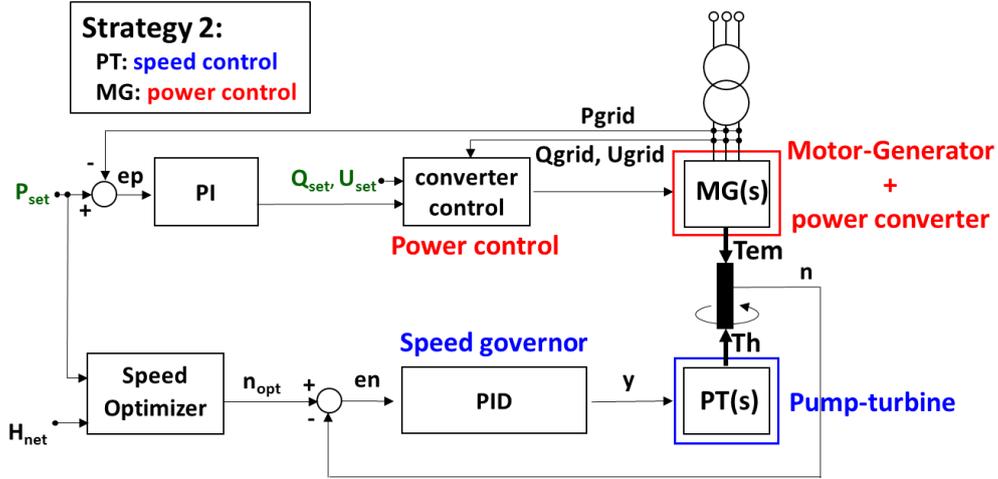

Fig. 5. Schematic block diagram of the variable speed unit controller in turbine mode, with control strategy 2 strategy 2: speed regulation with turbine and power regulation strategy with converter.

The 1D model includes the control system related to variable speed technologies described above and allows to simulate the provision of Frequency Containment Reserve (FCR), Synthetic Inertia (SI), Fast Frequency Response (FFR) and automatic Frequency Restauration Reserve (aFRR) ancillary services. The control system parameters have also been optimized to achieve high control performance. Since the rotational speed of the variable speed drives is decoupled from the power network frequency, the ancillary services are achieved by a change of the power setpoint sent to the unit controller. Fig. 6 illustrates the schematic block diagram of the active power set point $P_o$ modification due to power network frequency deviation $\Delta f_{grid}$. The aFRR service is achieved by the change of the nominal power setpoint $P_o$, which is independent of the grid frequency. The FCR branch generates a power setpoint change proportional to the grid frequency deviation with a first order transfer function defined by a time constant $\tau_P$ and a gain $K_p = -1/Bs$, where Bs is the permanent droop. The synthetic inertia response to the frequency RoCof is achieved by a first order transfer function defined by a time constant $\tau_d$ in series with a derivative term and a gain $K_d$ which is set to correspond to the mechanical time constant of the unit $\tau_m$, [7], [8]. The FFR ancillary service is represented with an independent branch, which activates a power step during a defined support duration when the frequency drops below a threshold. The results related to the evaluation of the FCR and FFR ancillary services capacity of Frades II PSPP can be found in [12], while this paper focuses on the evaluation of the Synthetic Inertia (SI) services.

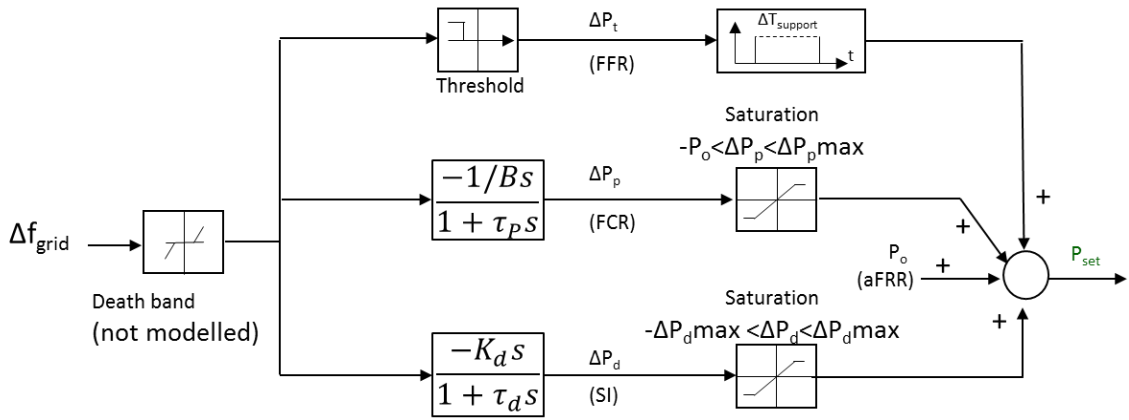

Fig. 6. Schematic block diagram of the active power set point modification due to frequency deviation with primary regulation and inertia emulation services.

The DFIM technology features some inherent speed deviation limits due to the maximum voltage amplitude of the 1st harmonic (slip frequency) in the rotor winding. With the active power of the DFIM converter being equal to the slip power, the allowable speed range is limited by the power capacity of the frequency converter. In the case of Frades 2, the speed of the unit should remain within the steady speed range 350-381 min$^{-1}$. Outside this speed range, the power electronics cannot maintain the speed of the unit for long without overheating the electrical components. It is nevertheless accepted that the unit makes speed excursions of ± 10 min$^{-1}$ outside the steady speed range during transients, [9], which set the transient speed range between 340-391 min$^{-1}$. During the normal generating mode, the runner can be operated at any speed within the authorized range. One strategy is to select the rotational speed $n_{opt}$ which maximizes the turbine efficiency for a given power. The corresponding allowable speed range as a function of the power are represented in the Fig. 7, with the steady speed range operation in green, the allowed transient speed in orange, and in blue the area delimiting the location of $n_{opt}$, which is bounded by the minimum and maximum head. It can be observed that, due to the admissible speed limitation, the optimal speed is equal to the minimum speed $n_{min}$ = 350 min$^{-1}$ over almost the entire power range, i.e. $n_{opt} = n_{min}$ for 0<P<0.98 pu. Consequently, the available stored kinetic energy is limited when operating at $n_{opt} = n_{min}$, compared to a higher rotational speed, as illustrated by the arrows representing the allowed transient speed drop in Fig. 7. To maximize control performances for SI and FCR services for negative and positive power network frequency deviations, the rotational speed setpoint has to be set to the middle of the continuous speed range $n_{middle}$ in order to maximize the margins prior to reach the upper or the lower speed limits, [12].

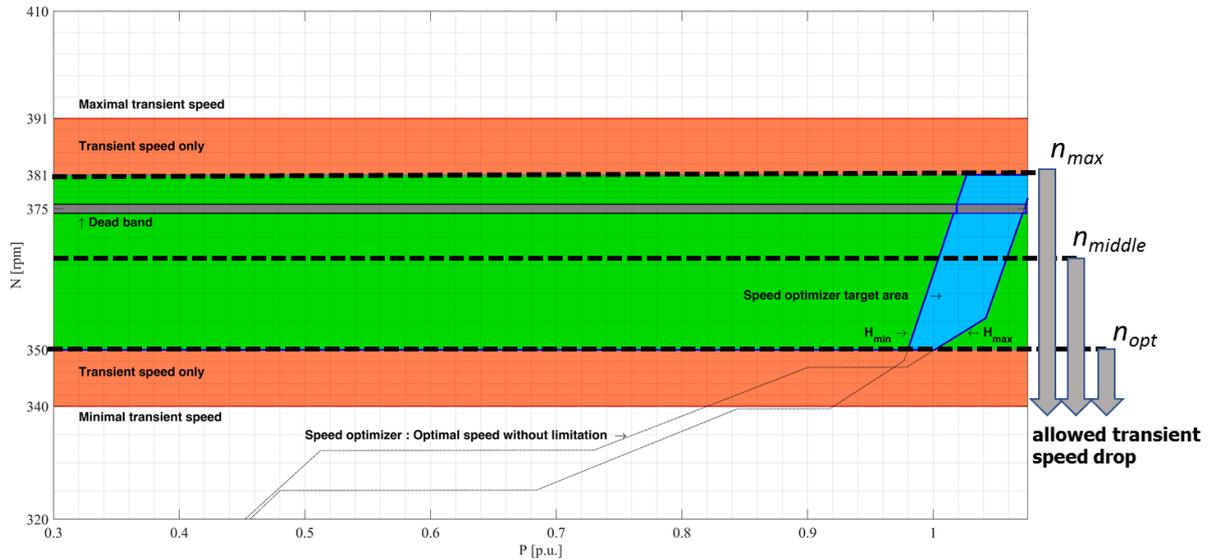

Fig. 7. Frades 2 allowable speed range as function of the power with the best efficiency speed marked in blue. The arrows illustrate the maximum allowed transient speed drop as function of the initial rotational speed of the unit.

## 4 Evaluation of synchronous and synthetic inertia by means of 1D numerical simulations

### 4.1 Infinite electrical power system

Numerical simulations have been performed with infinite power network with both fixed speed and variable speed motor-generator SIMSEN 1D simulation models of 1 unit of Frades 2 PSPP as presented in Fig. 8. These simulations aim to i) demonstrate the validity of equation (7) to predict active power injection or abortion in the power network resulting from grid frequency deviation for different *RoCoF* and ii) to demonstrate the ability of DFIM motor-generators to inject or absorb similar active power when synthetic inertia control system is included in the unit controller. As shown in Fig. 8 left, the fixed speed simulation includes a synchronous motor-generator for Frades 2 ($S_n$ = 420 MVA, $N_n$ = 375 min$^{-1}$, $U_n$ = 15.5 kV, $\tau_m$ = 7.9 s), a constant voltage excitation, a transformer ($x_{cc}$=12%) and a connection to an infinite grid. The simulation model of the synchronous machine takes into account sub-transient characteristics, see [11], using realistic parameters of the electrical equivalent scheme. Fig. 9, Fig. 10, and Fig. 11 present numerical simulation results for the case where the power network grid frequency is following a linear variation down and up with *RocoF* values of respectively ±0.5 Hz/s, 1 Hz/s and 2 Hz/s. In these figures, $P_{unit}$, $n$ and $f_{grid}$ correspond respectively to the output active power of the synchronous motor-generator, the unit rotational speed and the grid frequency. The simulation results show that the synchronous motor-generator is successively injecting and absorbing active power of average magnitude corresponding to 31 MW, 63 MW and 127 MW which is in line with the predictions of Table 1 with a mechanical time constant of $\tau_m$ = 7.9 s, which confirms the validity of equation (7). However, it could be noticed that the active power time evolution is the superposition from the mean active power injection or absorption due to rotating masses kinetic energy variation, and the rotor electromechanical oscillating mode, [11], leading to active power oscillations characterized by a natural frequency of about 1 Hz which damps out rather quickly.

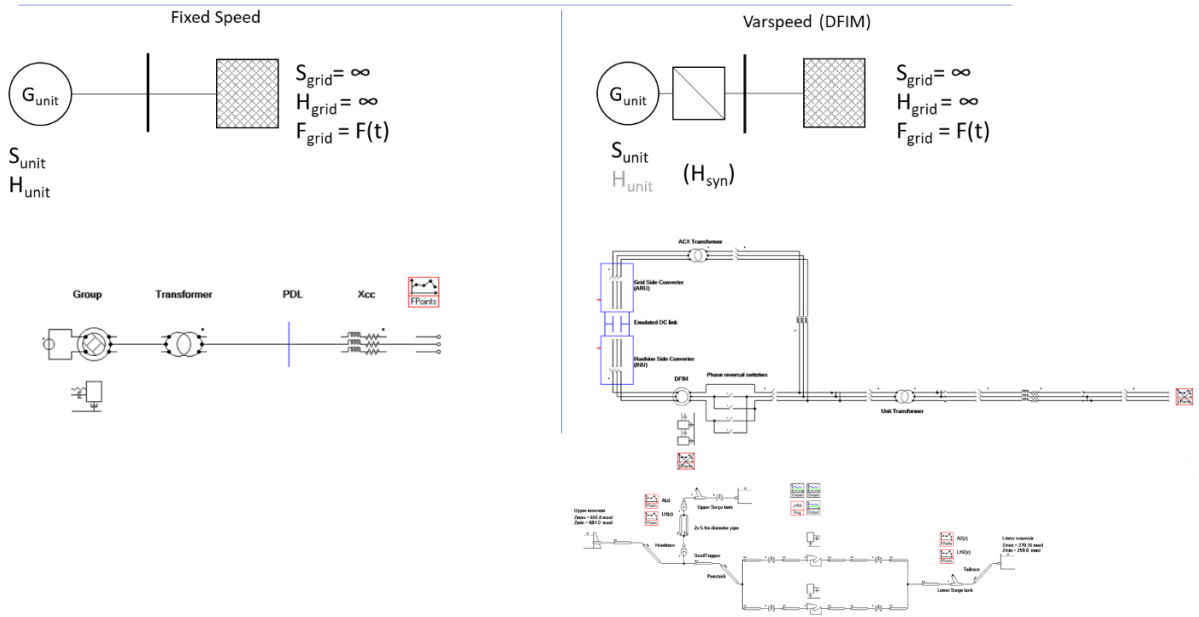

Fig. 8. Detailed electrical system and related 1D SIMSEN simulation models of Frades 2 for fixed speed (left) and variable speed (right) with infinite power network and related Frades 2 hydraulic system SIMSEN model used for the variable speed simulations.

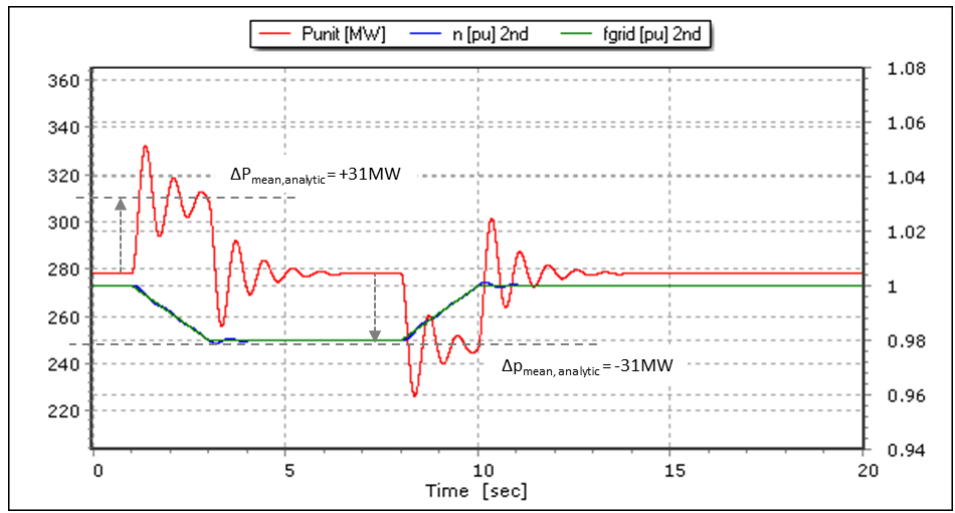

Fig. 9. Synchronous inertia power output with infinite grid and RoCoF of 0.5 Hz/s during 2 s.

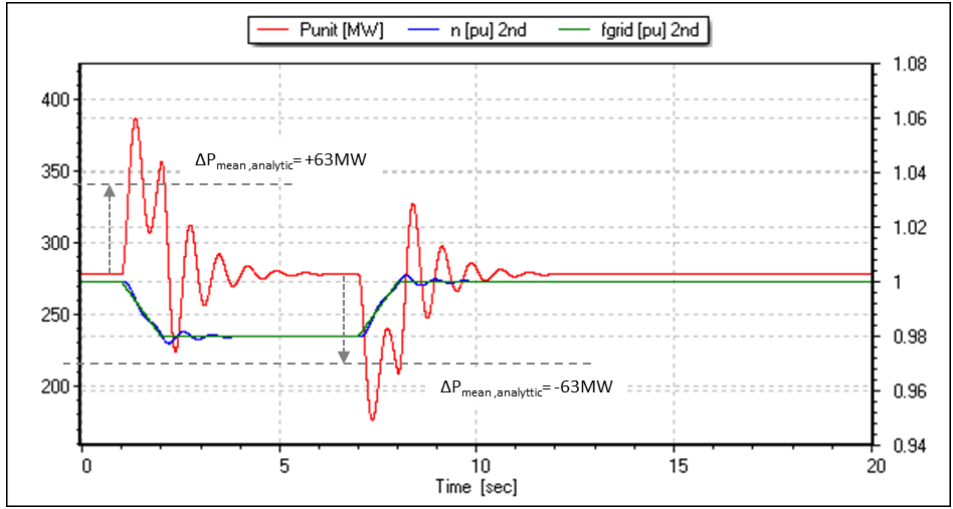

Fig. 10. Synchronous inertia power output with infinite grid and a RoCoF of 1 Hz/s during 1 s.

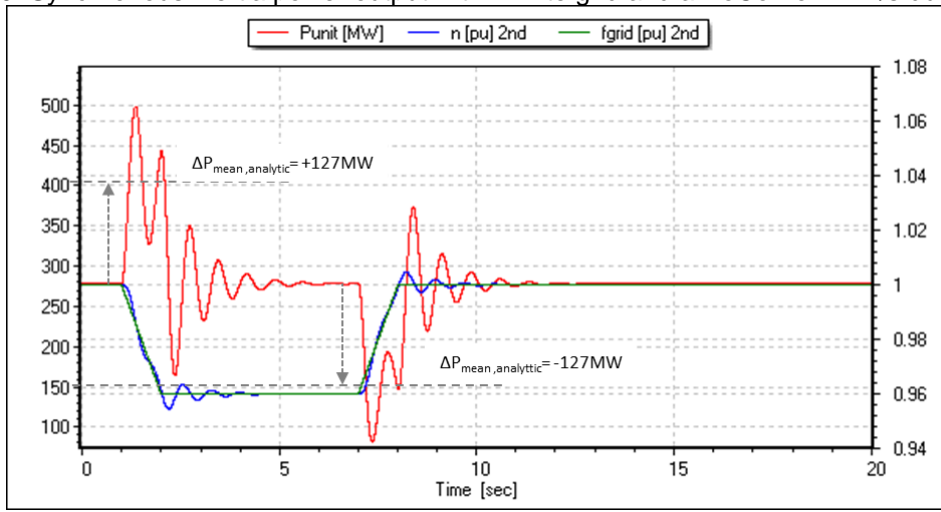

Fig. 11. Synchronous inertia power output with infinite grid and a RoCoF of 2 Hz/s during 1 s.

As shown in Fig. 8 right, the variable speed simulation includes the DFIM motor-generator and rotor-cascade including VSI power converter of Frades 2 ($S_n$ = 420 MVA, $N_n$ = 350 - 381 min$^{-1}$, $U_n$ = 15.5 kV, $\tau_m$ = 7.9 s) with the related converter control on the machine and grid side, a transformer ($x_{cc}$ = 12%) and a connection to an infinite grid. The simulation model of the DFIM takes into account transient characteristics and RMS model of the VSI rotor cascade, see [16], [17], [18], [24]. The simulations are performed without FCR and FFR contribution and only with the synthetic inertia branch (SI) of the unit controller shown in Fig. 6 considering gain values of $K_d$ = 4 or 8 s and a time constant of the filter of $\tau_d$ = 0.1 s. Fig. 12 and Fig. 13 present numerical simulation results for the case where the power network grid frequency is following a linear variation down and up for *RocoF* values of respectively 1 Hz/s and 2 Hz/s with $K_d$ = 8 s ≈ $\tau_m$. In these figures, $P_{unit}$, $n$ and $f_{grid}$ correspond respectively to the output active power of the DFIM motor-generator, the unit rotational speed and the grid frequency, while $h_1$, $q_1$, $t_1$, $n_1$ and $y_1$ correspond respectively to the pump-turbine of Unit 1 net head, discharge, torque, rotational speed and guide vane opening. The simulation results show that the motor-generator is successively injecting and absorbing active power of average magnitude corresponding to 64 MW and 128 MW which corresponds to the average values obtained with the synchronous machine without the presence of the electromechanical mode oscillations. The results are also in line with the predictions of Table 1. Fig. 14 presents numerical simulation results for the case where the power network grid frequency is following a linear variation down and up for Rocof values of 2 Hz/s with $K_d = 4$ s $\approx \tau_m/2$. As expected, since the gain of the synthetic inertia branch is divided by a factor two, the active power injection and absorption is also divided by a factor of 2 from 128 MW down to 64 MW, evidencing the linear relation between the control parameter $K_d$ and the magnitude of active power variation during constant *RoCoF* according to equation (5).

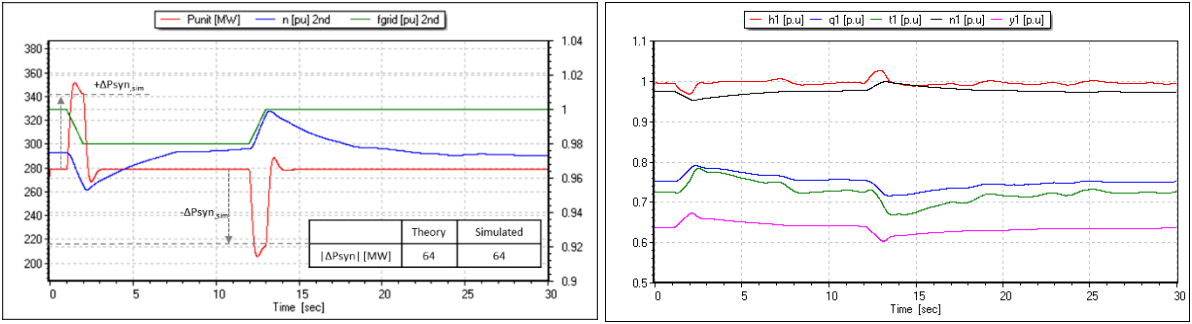

Fig. 12. Synthetic inertia power output with infinite grid and a RoCoF of 1 Hz/s for a duration of 1 s and Kd = 8 s and $\tau$d = 0.1 s.

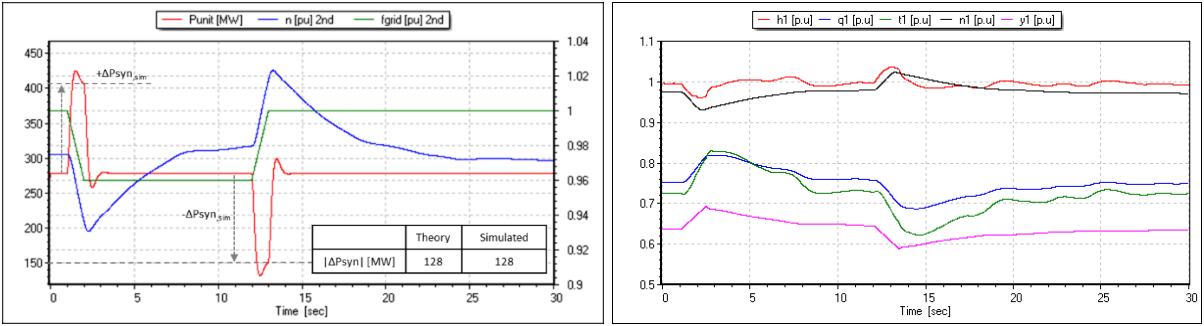

Fig. 13. Synthetic inertia power output with infinite grid and a RoCoF of 2 Hz/s for a duration of 1 s and Kd = 8 s and $\tau$d = 0.1 s.

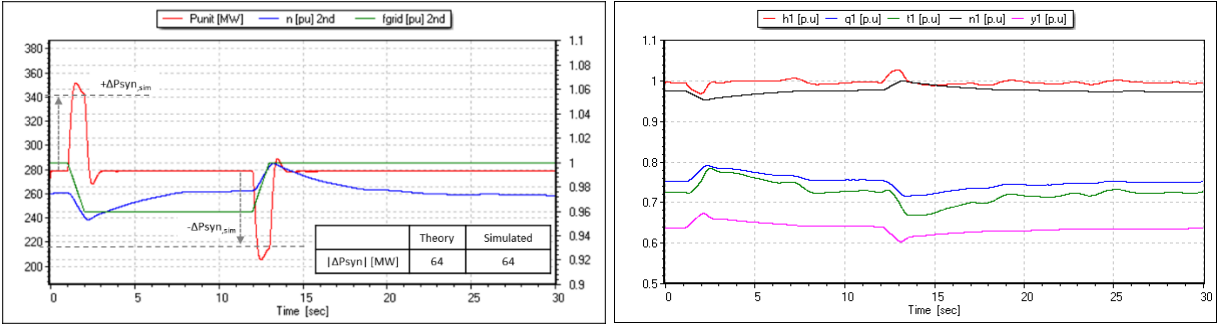

Fig. 14. Synthetic inertia power output with infinite grid and a RoCoF of 2 Hz/s for a duration of 1 s and Kd = 4 s and $\tau$d = 0.1 s.

## 4.2 Islanded electrical power system

Fig. 15 presents the power network configuration considered for the simulations performed in the case of islanded electrical power network with 1 unit of Frades 2 PSPP of 395 MW and a power network of 4.4 GW modelled with an equivalent synchronous generator. The electrical equivalent scheme and control parameters of the grid equivalent unit are taken from the Form 10 of the French grid code of RTE, [23], considering voltage control as well as permanent droop of BS = 10%. The mechanical time constant of this equivalent generator is $\tau_m$ = 7.9 s so that this generator is ~10 times larger than the Frades 2 motor-generator.

A load change is applied to obtain a power contribution from the hydro unit and from the grid equivalent unit, which is due to either the synthetic inertia or the synchronous inertia. The same load change of $\Delta P_{load}$ = +/-20% is applied to achieve a target *RoCoF* of 1 Hz/s over about 1 s. For the variable speed machine, simulations have been performed with three different values of synthetic inertia gain of $K_d$ = 0 s, 4 s and 8 s with a time constant of the filter of $\tau_d$ = 0.1 s. Fig. 16 presents the simulation results obtained for load change of $\Delta P_{load}$ = -20% while the Fig. 17 presents the simulations results obtained for a load change of $\Delta P_{load}$ = +20%. As expected, the active power absorption or injection of the variable speed unit is proportional to the synthetic inertia gain value $K_d$, and corresponds to zero when $K_d$ = 0 s, while it matches remarkably well the synchronous motor-generator active power in the case $K_d$ = 8 s ≈ $\tau_m$. It is important to mention that the grid frequency, here taken from the rotational speed of the grid equivalent unit, are perfectly superposed for the case $K_d$ = 8 s ≈ $\tau_m$. These simulation results are demonstrating that variable speed units are able to provide the same inertia contribution in islanded and infinite power networks, provided it includes unit controller with synthetic inertia contribution set with $K_d$ = $\tau_m$. Moreover, it requires sufficient rotational speed margin prior to reach maximum or minimum transient speed in order to allow the unit to inject or absorb the full active power contribution, and that contribution remains within allowable transient active power limits, *i.e.* voltage and current limits of the motor-generator and related VSI rotor cascade. The filter time constant of the synthetic inertia branch is so that the response time is below 500ms. This implies of course that the underlying converter control is faster than this filter time constant.

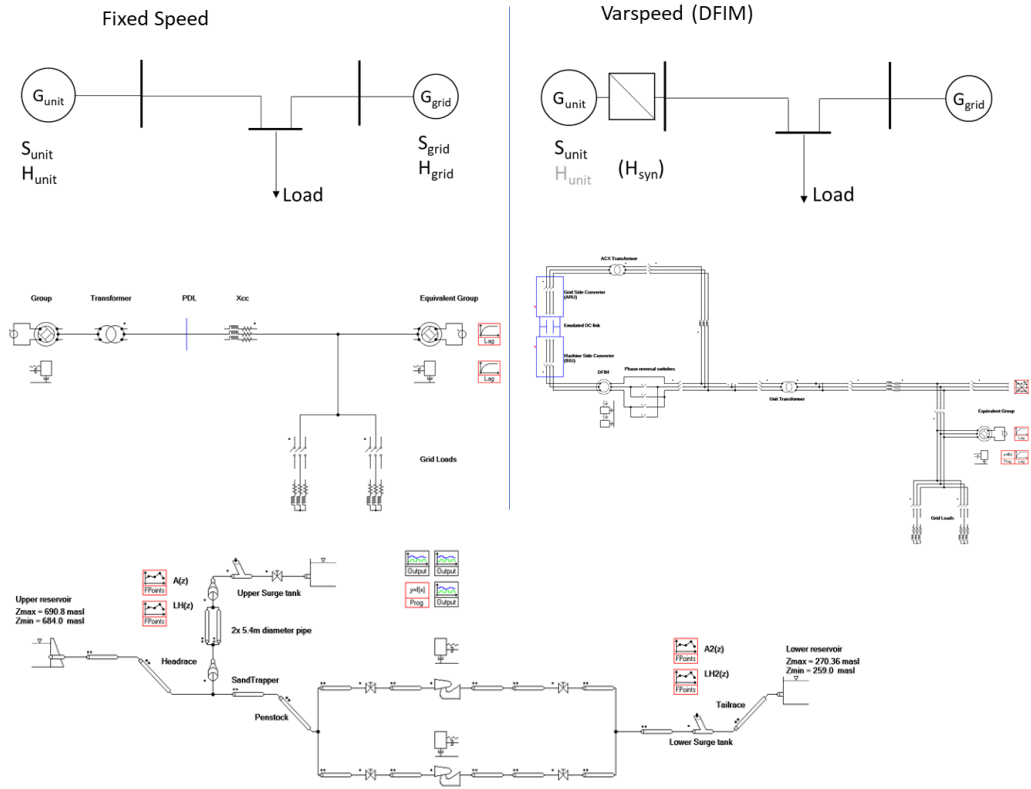

Fig. 15. Detailed electrical system and related 1D SIMSEN simulation models of Frades 2 for fixed speed (top left) and variable speed (top right) with islanded power network of 4.4 GW and related Frades 2 hydraulic system SIMSEN model (bottom).

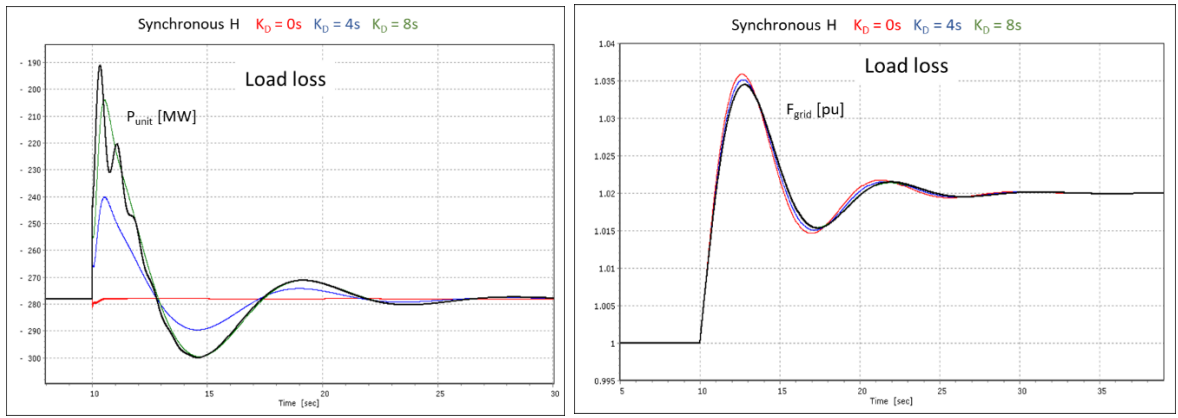

Fig. 16. Comparison of power output between synchronous and synthetic inertia, after load loss, with and a target RoCoF of 1 Hz/s, for several values of $K_d$.

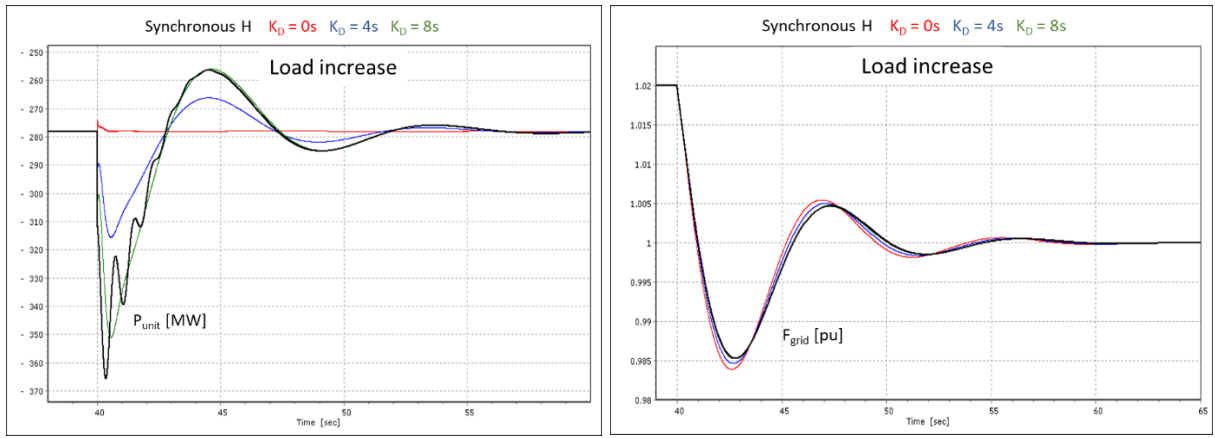

Fig. 17. Comparison of power output between synchronous and synthetic inertia, after load increase, with and a target RoCoF of 1 Hz/s, for several values of $K_d$.

## 4.3 Discussions

The goal the present simulations is simply to demonstrate that a variable speed unit with synthetic inertia can correctly support grid frequency deviation in the moments after a power unbalance. This takes advantage of the real kinetic energy stored in the rotor. Synthetic inertia could even expose a higher inertia constant to the grid as the physical inertia of the rotor. The consequence is simply a wider speed deviation of the rotor which can be supported by variable speed drive. The maximal speed deviation depends on technology, on waterways and on turbine governor performances.

Power electronic interfaced generation units, like the variable speed hydro power plant presented in this paper, have a control algorithm that fully relies on the presence of synchronous generation units in the grid. These algorithms control the injected power and are categorized as 'grid following' algorithm. In a grid featuring low synchronous generation means, it becomes difficult, if not impossible, to operate such 'grid following' controllers. Consequently, the stability of electrical power network featuring low rotational inertia level remains a challenging task, and requires appropriate control strategies consisting in a mix of 'grid following' and 'grid forming' capabilities, where the 'grid forming' capabilities attempt to emulate other fundamental properties of synchronous generation means, as presented in [25] [26].

## 5 Conclusion

The evaluation of synchronous and synthetic inertia contributions required a special attention since there is no grid code nor metrics available for these ancillary services. An analytical expression was derived to quantify the magnitude of the active power that a synchronous machine would inject or absorb in case a large power network frequency deviation would occur. This simple analytical expression, which was validated with detailed time domain simulations, enables to show that the magnitude of active power injected or absorbed is proportional to the hydro unit mechanical time constant $\tau_m$ for a given Rate of Change of the frequency (*RoCoF*). The analytical approach also shows that units operating in turbine or in pump contribute both to synchronous inertia since both feature rotating masses storing angular kinetic energy similarly to a flywheel. It also shows that, for hydraulic short circuit operation involving separated pump and turbine, as it is the case in Frades 2, the contribution to synchronous inertia is the sum of the pump and the turbine unit's synchronous inertia contributions.

Unlike synchronous machines, the rotational speed of variable speed units is disconnected from the power network frequency. Therefore, inertial response can be emulated by variable speed unit, using appropriate control structure to provide so-called synthetic inertia to the power network. Inertia emulation control structure is based on the swing equation to achieve active power injection or absorption which is also proportional to the *RoCof* like for a synchronous machine, and thus replicate the corresponding flywheel effect. To illustrate this, a specific grid stability study was conducted by means of 1D SIMSEN simulation for the demonstrator of Frades 2 with both fixed speed and variable speed technologies in case of operation connected to an infinite power network or an islanded synchronous power network of 4.4 GW. The time domain simulations performed with full hydroelectric model of Frades 2 including the detailed waterway, pump-turbine and motor-generator with the related control system, enabled to show that variable speed units inject or absorb the same active power as a synchronous machine in case of frequency deviation, provided that the power network features sufficient amount of synchronous inertia so that the variable speed unit can be operated in grid following mode. In case of islanded or even isolated power network with small or inexistent synchronous inertia, the controller of the variable speed unit shall allow for grid forming capability, while grid stability will remain a challenging task.

## Acknowledgments

The Hydropower Extending Power System Flexibility (XFLEX HYDRO) project has received funding from the European Union's Horizon 2020 research and innovation programme under grant agreement No 857832.

**Author(s)**

**Dr Christophe Nicolet** (christophe.nicolet@powervision-eng.ch) graduated from the Ecole polytechnique fédérale de Lausanne, EPFL, in Switzerland, and received his Master degree in Mechanical Engineering in 2001. He obtained his PhD in 2007 from the EPFL Laboratory for Hydraulic Machines. Since, he is managing director and principal consultant of Power Vision Engineering Sàrl in St-Sulpice, Switzerland, a company addressing optimization of hydropower transients and operation. He is also external lecturer at EPFL in the field of "Transient Flow".

**Dr Antoine Béguin** (antoine.béguin@powervision-eng.ch) is a project engineer at Power Vision Engineering Sàrl in St-Sulpice, Switzerland. He graduated from the Ecole polytechnique fédérale de Lausanne, EPFL, in Switzerland, and received his Master degree in electrical engineering in 2006. He obtained his PhD in 2011 from the same institution in the Laboratory for Power Electronics. Since 2011, he is working with Power Vision Engineering Sàrl in St-Sulpice, Switzerland, on transient phenomena, simulation and analysis of the dynamic behavior of hydroelectric power plants and their interactions with the power network. As a software development and communication specialist, he is responsible for the development of the Hydro-Clone system.

**Dr Matthieu Dreyer** (matthieu.dreyer@powervision-eng.ch) is a project engineer at Power Vision Engineering Sàrl in St-Sulpice, Switzerland. He graduated from the Ecole polytechnique fédérale de Lausanne, EPFL, in Switzerland, and received his Master degree in Mechanical Engineering in 2010. He obtained his PhD in 2015 from the same institution in the Laboratory for Hydraulic Machines. After two years of post-doctoral research in the same laboratory, he joined Power Vision Engineering Sàrl in 2017. Since then, he is responsible for the integration of Hydro-Clone system in hydropower plants.

**Dr Sébastien Alligné** (sebasien.alligne@powervision-eng.ch) is a project engineer at Power Vision Engineering Sàrl in St-Sulpice, Switzerland. He graduated in Mechanical Engineering at the Ecole Nationale Supérieurs d'Hydraulique et de Mécanique de Grenoble, ENSHMG, in 2002. He then worked in automotive and hydraulic industries between 2002 and 2007. He obtained his PhD degree at EPFL Laboratory for Hydraulic Machines in the field of simulation of complex 3D flow in hydraulic machinery in 2011. Since 2011, he is responsible of hydraulic system dynamic behaviour modelling and of CFD computations at Power Vision Engineering.

**Dr Alexander Jung** (alexander.jung@voith.com) studied mechanical engineering at University of Stuttgart and Northwestern University and graduated in 1993. He holds a doctoral degree and has about 30 years of experience in method development, gained by working for, or chairing, every physical discipline relevant to hydropower. He joined Voith in 2004 and worked several years as a hydraulic designer for large Francis and pump turbine projects before he took over various leadership positions. Since April 2019 he is responsible for Digital Hydro Technology at Voith Hydro.

**Diogo Cordeiro** (Diogo.Cordeiro@edp.com) is a Senior Innovation Project Manager at EDP Gestão da Produção de Energia SA (EDPP), a Portuguese conventional Generation Utility that manage a portfolio of a more than 50 hydropowerplants, 3 thermal powerplants and is a leading actor of floating solar technology. Diogo is the coordinator of the Frades 2 demonstrator, and EDPP PMO, in the scope of the XFLEX HYDRO project.  With 15 years' experience in the Energy sector as a project manager for continuous improvement programs, innovation/technology, processes, and training management, namely at EDP Lean Program, EDP Corporate University and EU Funding Innovation / Startup Engagement Programs. Before XFLEX HYDRO project Diogo was head of corporate Lean Program for all EDP Group and training manager at EDP University. Diogo holds an MBA from the Lisbon MBA, and an MSc degree in Electrotechnical Engineering from Instituto Superior Técnico, Portugal.

**Prof. Carlos Moreira** (carlos.moreira@inesctec.pt) graduated in Electrical Engineering in the Faculty of Engineering of the University of Porto - FEUP (2003) and completed his PhD in Power Systems in November 2008, also from the University of Porto. He is a Senior Researcher in the Centre for Power and Energy Systems of INESC TEC since September 2003. In February 2009 he joined the Department of Electrical Engineering of FEUP as Assistant Professor. He lectures several classes in graduation and MSc courses in Electrical Engineering and Power Systems and supervises the research activities of several MSc and PhD students. His main research interests are related to microgrids operation and control, dynamics and stability analysis of electric power systems with increasing shares of converter interfaced generation systems and grid code development.